\documentclass[conference]{IEEEtran}
\IEEEoverridecommandlockouts
\usepackage{cite}
\usepackage{amsmath,amssymb,amsfonts}
\usepackage{algorithmic}
\usepackage{gensymb}
\usepackage{graphicx}
\usepackage{textcomp}
\usepackage{xcolor}
\usepackage{academicons}
\usepackage{orcidlink}
\usepackage{algorithmic}
\usepackage{textcomp}
\usepackage{gensymb}
\usepackage{xcolor}
\def\BibTeX{{\rm B\kern-.05em{\sc i\kern-.025em b}\kern-.08em
    T\kern-.1667em\lower.7ex\hbox{E}\kern-.125emX}}

\usepackage{cite}         
\usepackage{hyperref}     
\hypersetup{
    colorlinks=true,
    linkcolor=blue,
    citecolor=blue,       
    filecolor=magenta,      
    urlcolor=blue,
    pdftitle={MACE},
    pdfpagemode=FullScreen,
} 

\usepackage{bm}
\usepackage{mathtools}
\usepackage{svg}
\usepackage{academicons}

\graphicspath{{fig/}}

\newcommand{\bmr}[1]{\bm{\mathrm{#1}}}

\newcommand{\nc}[1]{\mathcal{N}_{\mathbb{C}}(#1)}
\newcommand{\Expectation}[1]{\mathbb{E}\left\{#1\right\}}
\newcommand{\expectation}[1]{\mathbb{E}\{#1\}}
\newcommand{\norm}[1]{\lVert#1\rVert}

\newcommand{\parenth}[1]{\left(#1\right)}

\newcommand{\C}{\mathbb{C}}
\newcommand{\Vdiag}{\bmr{V}_{l,k}^{\text{diag}}[b]}

\graphicspath{{figs/}}

\definecolor{orcidlogocol}{HTML}{A6CE39}
\def\BibTeX{{\rm B\kern-.05em{\sc i\kern-.025em b}\kern-.08em
    T\kern-.1667em\lower.7ex\hbox{E}\kern-.125emX}}

\begin{document}

\title{Master-Assisted Channel Estimation for Cell-Free Massive MIMO Networks
\thanks{This research was carried out at the ESAT Laboratory of KU Leuven, in the frame of Research Project FWO nr.~G0C0623N ``User-centric distributed signal processing algorithms for next generation cell-free massive MIMO based wireless communication networks''. The scientific responsibility is assumed by its authors. \\ 
All simulation results can be reproduced using Matlab code available at: \href{https://github.com/AndreasAgg/master-assisted-CE.git}{https://github.com/AndreasAgg/master-assisted-CE.git}.}
}

\author{Andreas Angelou~\orcidlink{0009-0004-6728-0753}, and Marc Moonen~\orcidlink{0000-0003-4461-0073} \\
\IEEEauthorblockA{\textit{Department of Electrical Engineering (ESAT)
} \\
\textit{STADIUS Center for Dynamical Systems, Signal Processing and Data Analytics}\\
\textit{KU Leuven} \\
Leuven, Belgium \\
\{andreas.angelou, marc.moonen\}@kuleuven.be}
}

\maketitle

\begin{abstract}
Cell-free massive multiple-input–multiple-output (CFmMIMO) is a key enabler for sixth-generation (6G) wireless communication networks, where distributed access points (APs) jointly serve user equipments (UEs). In commonly adopted channel models for CFmMIMO networks, inter-AP channel correlation is assumed to be absent, thereby eliminating the potential benefits of centralized processing. However, by carefully designing the pilot transmission phase, the AP received signals during pilot transmission can become correlated, and thus, centralization can improve channel estimation performance, despite the absence of inter-AP channel correlation. In this paper, we propose a channel estimation scheme, termed master-assisted channel estimation (MACE), that aims to leverage inter-AP signal correlation by means of partially centralized processing and hence improve channel estimation performance. In MACE, a subset of APs fuse and forward their received pilot signals to a master AP, which then performs channel estimation using the fused signals together with its locally received signals. This scheme strikes a balance between local and fully centralized processing by leveraging inter-AP signal correlation, while reducing fronthaul signaling and computational complexity. Numerical experiments demonstrate that MACE consistently outperforms local channel estimation, where inter-AP signal correlation is neglected.
\end{abstract}


\section{Introduction}
Cell-free massive multiple-input-multiple-output (CFmMIMO) is a form of Network MIMO \cite{gesbert2010multi}, in which a large number of access points (APs) are distributed across a geographic region to jointly serve user equipments (UEs). This distributed AP deployment enables CFmMIMO to provide macro-diversity, cancel interference, and offer uniform quality of service to all UEs in the network \cite{ngo2015cell}. 

During the pilot transmission phase, the APs typically perform channel estimation using minimum mean-squared error (MMSE)-based estimators. Linear MMSE (LMMSE) estimation is commonly adopted under Rayleigh and Rician fading \cite{bjornson2019making, wang2021uplink}, and several variants have been proposed, such as element-wise MMSE \cite{ning2025spectral} and phase-aware MMSE \cite{ozdogan2019performance}. 

In the standard pilot transmission phase design, where UEs are assigned deterministic pilot signals, and in the absence of inter-AP channel correlation, centralized processing does not offer performance gains for channel estimation \cite{demir2021foundations}. However, a slight modification to the pilot transmission phase design outlined in \cite{van2022gevd}, originally proposed in \cite{pourya}, results in correlated received signals across APs, showcasing clear potential for performance improvement with centralized processing. However, centralized processing incurs increased computational complexity and fronthaul signaling. These trade-offs motivate the development of channel estimation schemes that explicitly leverage inter-AP signal correlation and balance performance with resource requirements.

In this paper, we propose a partially centralized channel estimation scheme, termed master-assisted channel estimation (MACE), inspired by the master-assisted distributed uplink operation \cite{angelou2026master}. In MACE, each UE is assigned a master AP (MAP) and a set of additional serving APs (ASAPs), which fuse and forward their received signals to the MAP. The MAP then performs channel estimation using the fused signals together with its locally received signals. MACE requires additional computational complexity and fronthaul signaling compared to the local channel estimation scheme, but significantly reduces fronthaul signaling and computational complexity compared to the fully centralized channel estimation scheme. Numerical experiments demonstrate that MACE consistently outperforms local channel estimation and approximates centralized channel estimation performance.

The contributions of this paper are summarized as follows: 
$i$) we introduce a channel estimation procedure, outlined in \hyperref[sec:local-cntr]{Section \ref{sec:local-cntr}}, which relies on the estimation of the line-of-sight (LoS) component and the channel \textbf{covariance} matrices, different from \cite{van2022gevd}, \cite{pourya}, and \cite{van2024distributed}, which rely on the estimation of the channel \textbf{correlation} matrices, and
$ii$) we develop MACE, a channel estimation scheme that leverages inter-AP signal correlation with reduced resource requirement compared to centralized channel estimation.

\section{System Model}
We consider a CFmMIMO network with $L$ APs, each equipped with $N$ antennas, and $K$ single-antenna UEs. For simplicity, we adopt the canonical CFmMIMO model, where all APs jointly serve all UEs. Our analysis can be easily extended for user-centric CFmMIMO, where each UE is served by a subset of APs \cite{demir2021foundations}. The APs are connected to a central processing unit (CPU) through ideal fronthaul links. The channel between UE $k$ and AP $j$ in coherence block $b$ is denoted by $\bmr{h}_{j,k}[b] \in \C^{N}$ and the collective channel of UE $k$ to all APs in coherence block $b$ is denoted by $\bmr{h}_{k}[b]  = \begin{bmatrix}
    \bmr{h}_{1,k}^T[b] \ \dots \ \bmr{h}_{L,k}^T[b]
\end{bmatrix}^T \in \C^{LN}$. We consider that the channels are constant over coherence blocks of $\tau_c$ samples. Since we only focus on channel estimation, we assume that $\tau_p = \tau_c$ samples are used for channel estimation in all coherence blocks. We adopt the pilot transmission phase design outlined in \cite{pourya}, which eliminates the need for pilot assignment algorithms and introduces inter-AP signal correlation, as demonstrated in \hyperref[sec:cntr]{Section \ref{sec:cntr}}. Specifically, in each coherence block $b$, UE $k$ chooses at random one pilot signal $\phi_{t_k[b]} \in \C^{\tau_p}$ out of $\tau_p$ mutually orthogonal pilot signals, where $t_k[b]\in\{1, \dots, \tau_p\}$ denotes the pilot index chosen by UE $k$ in coherence block $b$, and $\norm{\bmr{\phi}_{t_k[b]}}^2 = \tau_p$. Furthermore, UE $k$ scales its pilot signal using a scaling factor $\gamma_k[b] \in \{-1,+1\}$ in coherence block $b$, where $\gamma_k[b]=-1$ with probability $1/2$ and $\gamma_k[b]=+1$ with probability $1/2$. Henceforth, the index $b$ denotes the coherence block index.

\subsection{Channel Model} \label{sec:channel-model}
We adopt the correlated Rician fading model, where the channel between UE $k$ and AP $j$ is given by
\begin{equation} \label{eq:local-channel}
    \bmr{h}_{j,k}[b] = \bar{\bmr{h}}_{j,k} + \breve{\bmr{h}}_{j,k}[b] \, ,
\end{equation}
where $\bar{\bmr{h}}_{j,k}$ is the deterministic LoS component, and $\breve{\bmr{h}}_{j,k}[b]$ is the random zero-mean non-LoS (NLoS) component. The NLoS component is distributed as $\breve{\bmr{h}}_{j,k}[b] \sim \nc{\bmr{0}_N, \breve{\bmr{{R}}}_{j,k}}$, where $\breve{\bmr{{R}}}_{j,k}  = \expectation{\breve{\bmr{h}}_{j,k}[b] \breve{\bmr{h}}_{j,k}^H[b]} \in \C^{N \times N}$ is the correlation matrix. The correlation matrix of the channel $\bmr{R}_{j,k} = \expectation{\bmr{h}_{j,k}[b] \bmr{h}_{j,k}^H[b]} \in \C^{N \times N}$ is then given by 
\begin{equation} \label{eq:corr}
    \bmr{R}_{j,k} = \bar{\bmr{h}}_{j,k} \bar{\bmr{h}}_{j,k}^H + \breve{\bmr{R}}_{j,k}\, .
\end{equation}
We assume that the LoS component $\bar{\bmr{h}}_{j,k}$ and the correlation matrix $\breve{\bmr{R}}_{j,k}$ remain constant over a large number of coherence blocks, and hence do not depend on the coherence block index $b$, as shown in \eqref{eq:corr}. Furthermore, we assume that the NLoS components are independent across APs. The collective channel is then given by
\begin{equation} \label{eq:collective-channel}
    \bmr{h}_{k}[b] = \bar{\bmr{h}}_k + \breve{\bmr{h}}_k[b]\, ,
\end{equation}
where $\bar{\bmr{h}}_{k}= \begin{bmatrix}
    \bar{\bmr{h}}_{1,k}^T \ \dots \ \bar{\bmr{h}}_{L,k}^T
\end{bmatrix}^T$ is the LoS component and $\breve{\bmr{h}}_{k}[b]= \begin{bmatrix}
    \breve{\bmr{h}}_{1,k}^T[b] \ \dots \ \breve{\bmr{h}}_{L,k}^T[b]
\end{bmatrix}^T$ is the NLoS component. The correlation matrix of the NLoS component $\breve{\bmr{R}}_{k}= \expectation{ \breve{\bmr{h}}_{k}[b] \breve{\bmr{h}}_{k}^H[b] } \in \C^{LN \times LN}$ is a block diagonal matrix consisting of the local correlation matrices $\{\breve{\bmr{R}}_{j,k}\}_{j=1}^L$, i.e., $\breve{\bmr{R}}_k = \text{blkdiag}\{\breve{\bmr{R}}_{1,k}, \dots, \breve{\bmr{R}}_{L,k}\}$. Finally, the correlation matrix of the collective channel $\bmr{R}_{k}  = \expectation{\bmr{h}_{k}[b] \bmr{h}_{k}^H[b]}  \in \C^{LN \times LN}$ is given by
\begin{equation}
    \bmr{R}_k = \bar{\bmr{h}}_{k}\bar{\bmr{h}}_{k}^H + \breve{\bmr{R}}_k\, .
\end{equation}

\subsection{Local and Centralized Channel Estimation}
\label{sec:local-cntr}
During the pilot transmission phase, AP $j$ receives the pilot signals of the UEs and constitutes the received signal $\bmr{Y}_j[b] \in \C^{N \times \tau_p}$, given by
\begin{equation} \label{eq:rec}
    \bmr{Y}_{j}[b] = \sum_{i=1}^K \sqrt{p_i} \bmr{h}_{j,i}[b] \gamma_i[b]\bmr{\phi}_{t_i[b]}^T + \bmr{N}_j[b]\, ,
\end{equation}
where $p_i$ is the transmit power used by UE $i$, and $\bmr{N}_j[b]$ is the noise matrix with independently distributed $\mathcal{N}(0,\sigma^2)$ entries. The covariance matrix of $\bmr{Y}_j[b]$ is $\bmr{Q}_j  = \expectation{\bmr{Y}_j[b]\bmr{Y}_j^H[b]} \in \C^{N \times N}$, and is given by
\begin{equation} \label{eq:Q-local}
    \bmr{Q}_j =  \tau_p \sum_{i=1}^K p_i \bmr{R}_{j,i} + \tau_p \sigma^2 \bmr{I}_N\, 
\end{equation}
where we used the fact that $\expectation{\gamma_i[b] \gamma_{i^{\prime}}[b]} = 0$ for $i \neq i^{\prime}$, and 1 otherwise. For the channel estimate of UE $k$, AP $j$ correlates the received signal in \eqref{eq:rec} with $\gamma_k[b]\bmr{\phi}_{t_k[b]}/\sqrt{\tau_p}$ to obtain the despread signal $\bmr{y}_{j,t_k}[b] \in \mathbb{C}^{N}$, given by
\begin{equation} \label{eq:local-despread}
    \bmr{y}_{j,t_k}[b] = \sqrt{p_k \tau_p} \bmr{h}_{j,k}[b] + \sum_{i\neq k} \sqrt{p_i \tau_p} \bmr{h}_{j,i}[b] \delta_i[b] + \bmr{n}_{j,t_k}[b]\, ,
\end{equation}
where $\bmr{n}_{j,t_k}[b] \sim \nc{\bmr{0}_N, \sigma^2 \bmr{I}_N}$, and $\delta_i[b]$ is a random variable because of the random selection of the pilot signals by the UEs and the random scaling factor $\gamma_i[b]$ \cite{pourya}. The random variable $\delta_i[b]$ can be $0$ (if $\phi_{t_k[b]} \neq \phi_{t_i[b]}$), $1$ (if $\phi_{t_k[b]}=\phi_{t_i[b]}$ and $\gamma_k[b] = \gamma_i[b]$), or $-1$ (if $\phi_{t_k[b]}=\phi_{t_i[b]}$ and $\gamma_k[b] \neq \gamma_i[b]$), with probabilities $1-1/\tau_p$, $1/(2\tau_p)$, and $1/(2\tau_p)$, respectively, and its mean and variance are given by $\expectation{\delta_i}=0$ and $\expectation{\delta_i^2}=1/\tau_p$, respectively.

The LoS component can be estimated by the APs in each coherence block by exponential averaging of the despread signal in \eqref{eq:local-despread}, i.e.,
\begin{equation}
    \sqrt{p_k \tau_p}{\bar{\bmr{h}}}_{j,k} = \expectation{\bmr{y}_{j,t_k}[b]} \approx \eta \bar{\bmr{y}}_{j,t_k}^{\prime}[b-1] + (1 - \eta) \bmr{y}_{j,t_k}[b] \ \, ,
\end{equation}
where $\bar{\bmr{y}}_{j,t_k}^{\prime}[b-1]$ denotes the estimated mean of the despread signal in coherence block $b-1$, and $\eta \in (0,1)$ close to $1$. 

\subsubsection{\textbf{Local Channel Estimation}} \label{sec:local}
If local channel estimation is used, AP $j$ directly performs LMMSE channel estimation and obtains the channel estimate 
\begin{equation} \label{eq:local-channel-est}
    \widehat{\bmr{h}}_{j,k}^{\text{(l)}}[b] = \bar{\bmr{h}}_{j,k} +  \sqrt{p_k \tau_p} \breve{\bmr{R}}_{j,k} \bmr{Q}_{j,t_k}^{-1} (\bmr{y}_{j,t_k}[b] - \bar{\bmr{y}}_{j,t_k})\, ,    
\end{equation}
where $\bar{\bmr{y}}_{j,t_k} = \expectation{\bmr{y}_{j,t_k}[b]}=\sqrt{p_k \tau_p} \bar{\bmr{h}}_{j,k}$, and $ \bmr{Q}_{j,t_k} = \Expectation{(\bmr{y}_{j,t_k}[b]-\bar{\bmr{y}}_{j,t_k})(\bmr{y}_{j,t_k}[b]-\bar{\bmr{y}}_{j,t_k})^H} $,
which (using $\expectation{\delta_i^2}=1/\tau_p$) is given by
\begin{equation} \label{eq:Qtk-local}
    \bmr{Q}_{j,t_k} = p_k\tau_p \breve{\bmr{R}}_{j,k} + \sum_{i\neq k}p_i \bmr{R}_{j,i} + \sigma^2\bm{\mathrm{I}}_N\, .
\end{equation}
Observing \eqref{eq:Q-local} and \eqref{eq:Qtk-local} we can write
\begin{equation} \label{eq:R-est-local}
    \breve{\bmr{R}}_{j,k} = \frac{1}{p_k\tau_p(\tau_p-1)}\parenth{\tau_p \bmr{Q}_{j,t_k} + \tau_p p_k \bar{\bmr{h}}_{j,k}\bar{\bmr{h}}_{j,k}^H - \bmr{Q}_j}\, .
\end{equation}
Equation \eqref{eq:local-channel-est} can be used to estimate ${\bmr{h}}_{j,k}[b]$, when estimates of $\breve{\bmr{R}}_{j,k}$ and $\bmr{Q}_{j,t_k}$ are available. Equation \eqref{eq:R-est-local} can be used to estimate $\breve{\bmr{R}}_{j,k}$, when estimates of $\bmr{Q}_{j,t_k}$ in \eqref{eq:Qtk-local} and $\bmr{Q}_j$ in \eqref{eq:Q-local} are available. In practice, the covariance matrices in \eqref{eq:Q-local} and \eqref{eq:Qtk-local} can be estimated in every coherence block, as proposed in \cite{pourya}, or even less frequently, since the channel statistics vary much more slowly than the channel realization. The covariance matrices in \eqref{eq:Q-local} and \eqref{eq:Qtk-local} are then estimated by exponential averaging, i.e., 
\begin{equation} \label{eq:exp-avg-local}
    \begin{split}
        \widehat{\bmr{Q}}_{j}[b] = \eta & \widehat{\bmr{Q}}_{j}[b-1] + (1-\eta)\bmr{Y}_j[b]\bmr{Y}^H_j[b] \\[0.2cm]
        \widehat{\bmr{Q}}_{j,t_k}[b] =  \eta & \widehat{\bmr{Q}}_{j,t_k}[b-1] + \\ (1-\eta) &(\bmr{y}_{j,t_k}[b] - {\bar{\bmr{y}}}_{j,t_k}) (\bmr{y}_{j,t_k}[b] - {\bar{\bmr{y}}}_{j,t_k})^H \, ,
    \end{split}
\end{equation}
where the hat $\widehat{\cdot}$ denotes estimated matrices. Then, $\breve{\bmr{R}}_{j,k}$ is estimated in every coherence block as
\begin{equation}
    \widehat{\breve{\bmr{R}}}_{j,k}[b] = \frac{1}{p_k\tau_p(\tau_p-1)}\parenth{\tau_p \widehat{\bmr{Q}}_{j,t_k}[b] + \tau_p p_k \bar{\bmr{h}}_{j,k}\bar{\bmr{h}}_{j,k}^H - \widehat{\bmr{Q}}_j[b]}
\end{equation}
using \eqref{eq:R-est-local} and the two covariance matrices in \eqref{eq:exp-avg-local} \cite{pourya}.
The normalized mean squared error (NMSE) of the channel estimate is given by 
\begin{equation} \label{eq:local-NMSE}
    \text{NMSE}_{j,k}^{\text{(l)}}  = \frac {\expectation{\norm{\bmr{h}_{j,k}[b] - \widehat{\bmr{h}}_{j,k}^{\text{(l)}}[b]}^2}} {\text{tr}(\bmr{R}_{j,k})}.
\end{equation}

Local channel estimation incurs no fronthaul signaling, since there is no communication between the APs and the CPU. The computational complexity is $\mathcal{O}(N^3)$, due to the matrix inversion in \eqref{eq:local-channel-est}. 

\subsubsection{\textbf{Centralized Channel Estimation}} \label{sec:cntr}
In centralized channel estimation, the APs send their received signals in \eqref{eq:rec} to the CPU, which then has access to the matrix $\bmr{Y}[b] = \begin{bmatrix}
        \bmr{Y}_{1}^T[b] \ \dots \ \bmr{Y}_{L}^T[b]
    \end{bmatrix}^T \in \mathbb{C}^{LN\times \tau_p}$, given by
\begin{equation} \label{eq:rec-cntr}
    \bmr{Y}[b] = \sum_{i=1}^K \sqrt{p_i} \bmr{h}_i[b] \gamma_i[b] \phi_{t_i[b]}^T + \bmr{N}[b]\, ,
\end{equation}
where $\bmr{N}[b]=\begin{bmatrix}
    \bmr{N}_1^T[b] \ \dots \ \bmr{N}_L^T[b]
\end{bmatrix}^T$. The covariance matrix of $\bmr{Y}[b]$ is $\bmr{Q}  = \expectation{\bmr{Y}[b]\bmr{Y}^H[b]}  \in \C^{LN \times LN}$, and is given by
\begin{equation} \label{eq:Q-cntr}
    \bmr{Q} = \tau_p \sum_{i=1}^K p_i \bmr{R}_i + \tau_p \sigma^2 \bmr{I}_{LN}\, .
\end{equation}
After correlating the received signals in \eqref{eq:rec-cntr} with the normalized pilot of UE $k$, the CPU obtains the despread signal $\bmr{y}_{t_k}[b] \in \C^{LN}$ given by
\begin{equation} \label{eq:cntr-despread}
    \bmr{y}_{t_k}[b] = \sqrt{p_k \tau_p} \bmr{h}_{k}[b] + \sum_{i\neq k} \sqrt{p_i \tau_p} \bmr{h}_{i}[b] \delta_i[b] + \bmr{n}_{t_k}[b]\, ,
\end{equation}
where $\bmr{n}_{t_k}[b] = \begin{bmatrix}
    \bmr{n}_{1,t_k}^T[b] \ \dots \ \bmr{n}_{L,t_k}^T[b]
\end{bmatrix}^T$. The LoS component of the collective channel can be estimated by exponential averaging, i.e., 
\begin{equation}
    \sqrt{p_k \tau_p}{\bar{\bmr{h}}}_{k} = \expectation{\bmr{y}_{t_k}[b]} \approx \eta \bar{\bmr{y}}_{t_k}^{\prime}[b-1] + (1 - \eta) \bmr{y}_{t_k}[b] \ \, ,
\end{equation}
where $\bar{\bmr{y}}_{t_k}^{\prime}[b-1]$ denotes the estimated mean of the despread signal in coherence block $b-1$.
The CPU then performs LMMSE channel estimation to obtain the collective channel estimate $\widehat{\bmr{h}}_k^{\text{(c)}}[b]  \in \C^{LN}$, given by 
\begin{equation} \label{eq:cntr-channel-est}
    \widehat{\bmr{h}}_k^{\text{(c)}}[b] = \bar{\bmr{h}}_k +  \sqrt{p_k \tau_p} \breve{\bmr{R}}_k \bmr{Q}_{t_k}^{-1} (\bmr{y}_{t_k}[b] -\bar{\bmr{y}}_{t_k})\, ,
\end{equation}
where $\bar{\bmr{y}}_{t_k} = \expectation{\bmr{y}_{t_k}[b]}=\sqrt{p_k \tau_p} \bar{\bmr{h}}_{k}$, and 
$
    \bmr{Q}_{t_k} = \Expectation{(\bmr{y}_{t_k}[b]-\bar{\bmr{y}}_{t_k})(\bmr{y}_{t_k}[b]-\bar{\bmr{y}}_{t_k})^H}
$,
which is given by
\begin{equation} \label{eq:Qtk-cntr}
    \bmr{Q}_{t_k} = p_k\tau_p \breve{\bmr{R}}_k  + \sum_{i\neq k} p_i \bmr{R}_i +\sigma^2\bm{\mathrm{I}}_{LN}\, .    
\end{equation}
Note that $\bmr{Q}_{t_k}$ is not block diagonal, thanks to the correlation matrices $\{\bmr{R}_i\}_{i \neq k}$ in \eqref{eq:Qtk-cntr}, which are not block diagonal because of the LoS components of the collective channels. These correlation matrices appear in \eqref{eq:Qtk-cntr} thanks to the random variable $\delta_i$, arising from the random pilot selection and the random scaling factor $\gamma_i$. Hence, the despread signals across the APs are correlated, and centralized channel estimation is expected to provide performance improvements over local channel estimation. Observing \eqref{eq:Q-cntr} and \eqref{eq:Qtk-cntr}, we can write
\begin{equation} \label{eq:R-est-cntr}
    \breve{\bmr{R}}_{k} = \frac{1}{p_k\tau_p(\tau_p-1)}\parenth{\tau_p \bmr{Q}_{t_k} + \tau_p p_k \bar{\bmr{h}}_{k}\bar{\bmr{h}}_{k}^H - \bmr{Q}}\, ,
\end{equation}
which can be used to estimate $\breve{\bmr{R}}_{k}$, when estimates of $\bmr{Q}_{t_k}$ in \eqref{eq:Qtk-cntr} and $\bmr{Q}$ in \eqref{eq:Q-cntr} are available. These two covariance matrices can be estimated by exponential averaging, i.e.,
\begin{equation}
    \label{eq:exp-avg-cntr}
    \begin{split}
        \widehat{\bmr{Q}}[b] = \eta & \widehat{\bmr{Q}}[b-1] + (1-\eta)\bmr{Y}[b]\bmr{Y}^H[b] \\[0.2cm]
        \widehat{\bmr{Q}}_{t_k}[b] =  \eta & \widehat{\bmr{Q}}_{t_k}[b-1] + \\ & (1-\eta)(\bmr{y}_{t_k} [b] - {\bar{\bmr{y}}}_{t_k})(\bmr{y}_{t_k}[b] - {\bar{\bmr{y}}}_{t_k})^H \, .
    \end{split}
\end{equation}

We focus on the part of the collective channel estimate, given in \eqref{eq:cntr-channel-est}, that corresponds to AP $l$, i.e., 
\begin{equation} \label{eq:cntr-local-channel-est}
    \widehat{\bmr{h}}_{l,k}^{\text{(c)}} = \widehat{\bmr{h}}^{\text{(c)}}_k[(l-1)N+1:lN]\, ,
\end{equation}
where we have extracted the elements of $\widehat{\bmr{h}}_{k}^{\text{(c)}}$ that correspond to indices from $(l-1)N+1$ to $lN$. The NMSE of the channel estimate between UE $k$ and AP $l$ is then given by 
\begin{equation} \label{eq:cntr-NMSE}
    \text{NMSE}_{l,k}^{\text{(c)}}  = \frac {\expectation{\norm{\bmr{h}_{l,k}[b] - \widehat{\bmr{h}}_{l,k}^{\text{(c)}}[b]}^2}} {\text{tr}(\bmr{R}_{l,k})}.
\end{equation}

For one UE, centralized channel estimation requires the exchange of $\tau_pLN$ complex scalars through fronthaul links, due to the communication between the APs and the CPU. The computational complexity is $\mathcal{O}(L^3N^3)$, due to the matrix inversion in \eqref{eq:cntr-channel-est}.

\section{Master-Assisted Channel Estimation} \label{sec:master}
In this section, we describe MACE, the main contribution of this paper. In MACE, each UE is served by a MAP and a set of ASAPs. Since we focus on the channel estimation of UE $k$, we denote the MAP of UE $k$ by $l$. All other $L-1$ APs are assumed to be the ASAPs of UE $k$. In MACE, the ASAPs fuse their received signals in \eqref{eq:rec} using a fusion vector $\bmr{v}_{j,k}[b] \in \C^{N}$ and send these fused signals to MAP $l$, which has access to its own locally received signals $\bmr{Y}_{l}[b] \in \C^{N \times \tau_p}$ and the fused signals $\bmr{v}_{j,k}^H[b] \bmr{Y}_{j}[b] \in \C^{1 \times \tau_p}$. In particular, ASAPs use their local channel estimates in \eqref{eq:local-channel-est} as fusion vectors, i.e., $\bmr{v}_{j,k}[b]=\widehat{\bmr{h}}_{j,k}^{\text{(l)}}[b]$. Thus, MAP $l$ has access to $\widetilde{\bmr{Y}}_{l}[b] \in \mathbb{C}^{(N+L-1)\times \tau_p}$, given by
\begin{equation} \label{eq:rec-master}
    \begin{split}
    \widetilde{\bmr{Y}}_{l}[b] = \Bigr[ 
        \bmr{Y}_{1}^H[b] \bmr{v}_{1,k}[b] \ \
        \dots \ \
        \bmr{Y}_{l-1}^H[b] \bmr{v}_{l-1,k}[b] \ \
        \bmr{Y}_{l,t_k}[b] \\[0.1cm]
        \bmr{Y}_{l+1}^H[b] \bmr{v}_{l+1,k}[b] \ \
        \dots \ \
        \bmr{Y}_{L}^H[b] \bmr{v}_{L,k}[b] 
        \Bigr]^H \, ,
    \end{split}
\end{equation}
which can be written as 
\begin{equation} \label{eq:rec-master-2}
    \widetilde{\bmr{Y}}_{l}[b] = 
\sum_{i=1}^K \sqrt{p_i}\bmr{z}_{l,ki}[b] \gamma_i[b] \phi_{t_i[b]}^T + \widetilde{\bmr{N}}_{l,k}[b]\, ,
\end{equation}
where, $\widetilde{\bmr{N}}_{l,k}[b] = (\Vdiag)^H \bmr{N}[b]$, with 
  $\Vdiag \in \mathbb{C}^{(LN) \times (N+L-1)}$ given by
\begin{equation} \label{eq:Vdiag}
    \begin{split}
        \Vdiag = \text{blkdiag} \big{\{} \widehat{\bmr{h}}_{1,k}^{\text{(l)}}[b], \ \dots \ , \ \widehat{\bmr{h}}_{l-1,k}^{\text{(l)}}[b], & \ \bmr{I}_N, \\ \widehat{\bmr{h}}_{l+1,k}^{\text{(l)}}[b], \ \dots, & \ \widehat{\bmr{h}}_{L,k}^{\text{(l)}}[b] \big{\}}\, ,
    \end{split}
\end{equation}
and $\bmr{z}_{l,ki}[b]\in \C^{N+L-1}$ is the fused channel of UE $i$, given by
\begin{equation} \label{eq:fused-channel}
    \bmr{z}_{l,ki}[b] = \bar{\bmr{z}}_{l,ki}[b] + \breve{\bmr{z}}_{l,ki}[b]\, ,
\end{equation}
with $\bar{\bmr{z}}_{l,ki}[b] = ({\Vdiag})^H \bar{\bmr{h}}_i$ and $\breve{\bmr{z}}_{l,i}[b] = (\Vdiag)^H  \breve{\bmr{h}}_i[b]$. The correlation matrix of the fused channel in \eqref{eq:fused-channel} is ${\bmr{R}}_{l,ki} =  \expectation{{\bmr{z}}_{l,ki}[b]{\bmr{z}}_{l,ki}^H[b]} \in \C^{(N+L-1) \times (N+L-1)}$, which is given by
\begin{equation} \label{eq:fused-correlation}
    {\bmr{R}}_{l,ki} = \expectation{\bar{\bmr{z}}_{l,ki}[b] \bar{\bmr{z}}_{l,ki}^H[b]} + \breve{{\bmr{R}}}_{l,ki} \approx (\Vdiag)^H \bmr{R}_i \Vdiag \, ,
\end{equation}
where
\begin{equation} \label{eq:fused-covariance}
    \breve{{\bmr{R}}}_{l,ki}  =  \expectation{\breve{\bmr{z}}_{l,ki}[b]\breve{\bmr{z}}_{l,ki}^H[b]}  \approx  (\Vdiag)^H \breve{\bmr{R}}_i \Vdiag \, ,
\end{equation}
is the block diagonal correlation matrix of the NLoS component $\breve{\bmr{z}}_{l,ki}[b]$\footnote{We use the approximations in \eqref{eq:fused-correlation} and \eqref{eq:fused-covariance} to simplify the analysis and derive the remaining equations of this section, since deriving closed-form expressions for the correlation and covariance matrices is tedious and results in extensive formulas.}. 
The covariance matrix of $\widetilde{\bmr{Y}}_l[b]$ in \eqref{eq:rec-master-2} is $\widetilde{\bmr{Q}}_l  = \expectation{\widetilde{\bmr{Y}}_{l}[b]\widetilde{\bmr{Y}}_{l}^H[b]}  \in \C^{(N+L-1) \times (N+L-1)}$ and is given by 
\begin{equation} \label{eq:Q-master}
    \widetilde{\bmr{Q}}_l= \tau_p \sum_{i=1}^K p_i {\bmr{R}}_{l,ki} + \tau_p \sigma^2 \bmr{I}_{N+L-1} \, .
\end{equation}
MAP $l$ then correlates $\widetilde{\bmr{Y}}_l[b]$ with the normalized pilot of UE $k$ to obtain the despread signal $\widetilde{\bmr{y}}_{l,t_k}[b] \in \C^{N+L-1}$, given by
\begin{equation} \label{eq:master-despread}
    \widetilde{\bmr{y}}_{l,t_k}[b] = \sqrt{p_k \tau_p}\bmr{z}_{l,kk}[b] + \sum_{i\neq k}\sqrt{p_i \tau_p}\bmr{z}_{l,ki}[b]\delta_i[b] + \widetilde{\bmr{n}}_{l,t_k}[b]\, ,
\end{equation}
where $\widetilde{\bmr{n}}_{l,t_k}[b] = (\Vdiag)^H \bmr{n}_{t_k}[b]$. 
The LoS component of the fused channel can be estimated by exponential averaging, i.e.,
\begin{equation}
    \sqrt{p_k \tau_p}\bar{\bmr{z}}_{l,kk} = \expectation{{\widetilde{\bmr{y}}}_{l,t_k}[b]} \approx \eta \overline{\widetilde{\bmr{y}}}_{l,t_k}^{\prime}[b-1] + (1-\eta)\widetilde{\bmr{y}}_{l,t_k}[b]  \, ,
\end{equation}
where $\overline{\widetilde{\bmr{y}}}_{l,t_k}^{\prime}[b-1]$ denotes the estimated mean of the despread signal in coherence block $b-1$.
MAP $l$ then performs LMMSE channel estimation to obtain the estimate of the fused channel $\bmr{z}_{l,kk}[b]$ given by
\begin{equation} \label{eq:master-channel-est}
        \widehat{\bmr{z}}_{l,kk}[b] = \bar{\bmr{z}}_{l,kk} +  \sqrt{p_k \tau_p} \breve{{\bmr{R}}}_{l,kk} \widetilde{\bmr{Q}}_{l,t_k}^{-1} (\widetilde{\bmr{y}}_{l,t_k}[b] - \overline{\widetilde{\bmr{y}}}_{l,t_k}),
    \end{equation}
where $\overline{\widetilde{\bmr{y}}}_{l,t_k}=\expectation{{\widetilde{\bmr{y}}}_{l,t_k}[b]} = \sqrt{p_k\tau_p} \bar{\bmr{z}}_{l,kk}$, and 
$
    \widetilde{\bmr{Q}}_{l,t_k} = \expectation{(\widetilde{\bmr{y}}_{l,t_k}[b]-\overline{\widetilde{\bmr{y}}}_{l,t_k})(\widetilde{\bmr{y}}_{l,t_k}[b]-\overline{\widetilde{\bmr{y}}}_{l,t_k})^H}\, ,
$
which is given by
\begin{equation} \label{eq:Qtk-master}
     \widetilde{\bmr{Q}}_{l,t_k} = p_k\tau_p \breve{{\bmr{R}}}_{l,kk} + \sum_{i \neq k} p_i {\bmr{R}}_{l,ki} + \sigma^2 \bmr{I}_{N+L-1} \, .
\end{equation}
Observing \eqref{eq:Q-master} and \eqref{eq:Qtk-master} we can write
\begin{equation} \label{eq:R-est-master}
    \breve{\bmr{R}}_{l,kk} = \frac{1}{p_k\tau_p(\tau_p-1)}\parenth{\tau_p \widetilde{\bmr{Q}}_{l,t_k} + \tau_p p_k \bar{\bmr{z}}_{l,kk}\bar{\bmr{z}}_{l,kk}^H - \widetilde{\bmr{Q}}_l}\, ,
\end{equation}
which can be used to estimate $\breve{\bmr{R}}_{l,kk}$, when estimates of $\widetilde{\bmr{Q}}_{l,t_k}$ in \eqref{eq:Qtk-master} and $\widetilde{\bmr{Q}}_l$ in \eqref{eq:Q-master} are available. 
These two covariance matrices can be estimated by exponential averaging, i.e.,
\begin{equation}
    \label{eq:exp-avg-master}
    \begin{split}
        \widehat{\widetilde{\bmr{Q}}}_l[b] = \eta & \widehat{\widetilde{\bmr{Q}}}_l[b-1] + (1-\eta)\widetilde{\bmr{Y}}_l[b]\widetilde{\bmr{Y}}_l^H[b] \\[0.2cm]
        \widehat{\widetilde{\bmr{Q}}}_{l,t_k}[b] =  \eta & \widehat{\widetilde{\bmr{Q}}}_{l,t_k}[b-1] + \\ (1-\eta) & (\widetilde{\bmr{y}}_{l,t_k}[b] - \overline{\widetilde{\bmr{y}}}_{l,t_k})(\widetilde{\bmr{y}}_{l,t_k}[b] - \overline{\widetilde{\bmr{y}}}_{l,t_k})^H \, .
    \end{split}
\end{equation}
We again focus on the part of $\widehat{\bmr{z}}_{l,kk}[b]$ in \eqref{eq:master-channel-est} that corresponds to MAP $l$, i.e., 
\begin{equation}\label{eq:master-local-channel-est}
    \widehat{\bmr{h}}_{l,k}^{\text{(m)}} = \widehat{\bmr{z}}_{l,kk}[l:l+N-1]\, .
\end{equation}
where we have extracted the elements of $\widehat{\bmr{z}}_{l,kk}$ that correspond to indices from $l$ to $l+N-1$. The NMSE is then given by
\begin{equation} \label{eq:master-NMSE}
    \text{NMSE}_{l,k}^{\text{(m)}}  = \frac {\expectation{\norm{\bmr{h}_{l,k}[b] - \widehat{\bmr{h}}_{l,k}^{\text{(m)}}[b]}^2}} {\text{tr}(\bmr{R}_{l,k})}.
\end{equation}

For one UE, MACE requires the exchange of $\tau_p(L-1)$ complex scalars through fronthaul links, due to the communication of the ASAPs and the MAP. This reduces the fronthaul signaling compared to the centralized channel estimation by a factor of $LN/(L-1)$. The computational complexity of MACE is $\mathcal{O}\parenth{(N+L-1)^3}$, due to the matrix inversion in \eqref{eq:master-channel-est}, which is substantially lower than that of centralized channel estimation.

\section{Numerical Experiments}
In this section, we compare local and centralized channel estimation with MACE. In centralized channel estimation, the MAPs are assumed to act as the CPU. We generate $200$ network realizations of $K$ UEs and $L$ APs in a square simulation area of $50$ m$^2$ and use $p_k=p=100$, $\sigma=1$, $\eta = 0.999$. We report the median results across the $200$ realizations using $300$ coherence blocks per realization (from $b=5001:5300$, after an initialization batch of $5000$ coherence blocks), with estimated covariance matrices for simulations and true covariance matrices for theoretical results.

We consider communication over a $20$ MHz bandwidth with a noise figure of $7$ dB. The large-scale-fading (LSF) coefficient $\beta_{j,k}$ is modeled as 
$
    \beta_{j,k} = -30.18 - 26\log_{10}(d_{j,k}) + F_{j,k}
$, where $d_{j,k}$ is the distance between AP $j$ and UE $k$ in meters, and $F_{j,k} \sim \mathcal{N}(0,4^2)$. 
The covariance matrices of the channels are generated using $[\breve{\bmr{R}}_{j,k}]_{n,m} = \frac{\beta^{\text{NLoS}}_{j,k}}{\sqrt{2\pi }\sigma_\phi} \int_{-\infty}^{+\infty} e^{j \pi (n-m)\sin{(\theta_{j,k} + \delta)}} e^{-\delta^2/(2\sigma_\phi^2)} d\delta$, where $\delta \sim \nc{0,\sigma_\phi^2}$, $\sigma_\phi^2 = 15\degree$, and $\beta^{\text{NLoS}}_{j,k} = \beta_{j,k} / (\kappa_{j,k}+1)$, with $\kappa_{j,k} = 10^{1.3-0.003 d_{j,k}}$. The $n$-th element of the LoS component is generated as $[\bar{\bmr{h}}_{j,k}]_n = \sqrt{\beta_{j,k}^{\text{LoS}}} e^{j \pi (n-1) \sin(\theta_{j,k})}$, where $\theta_{j,k}$ is the angle of arrival to UE $k$ observed by AP $j$ and $\beta^{\text{LoS}}_{j,k} =  \beta_{j,k}\kappa_{j,k}/(\kappa_{j,k}+1)$. Each UE selects its MAP by identifying the AP with the largest LSF coefficient $\beta_{l,k}$.

\begin{figure}[!t]
  \centering
  \def\svgwidth{0.85\columnwidth}
  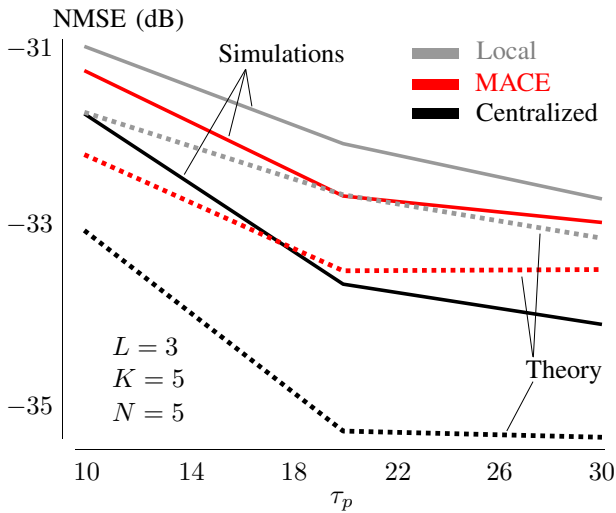
  \caption{NMSE of all estimation schemes for varying $\tau_p$. MACE and centralized channel estimation outperform local channel estimation.}
  \label{fig:tau}
\end{figure}

\begin{figure}[!t]
  \centering
  \def\svgwidth{0.85\columnwidth}
  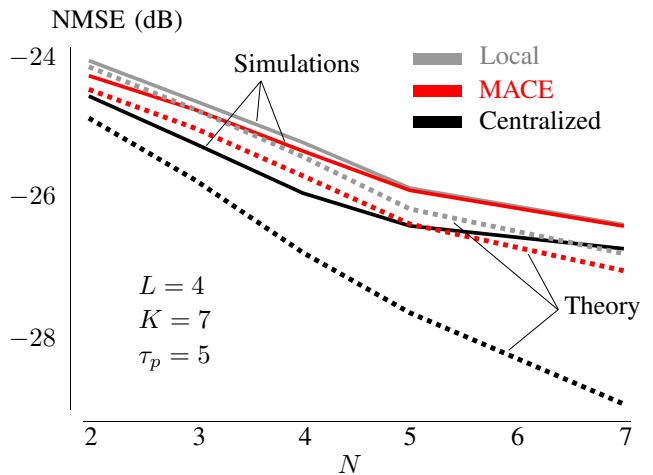
  \caption{NMSE of all estimation schemes for varying $N$. MACE approaches the performance of local channel estimation as $N$ increases.}
  \label{fig:antennas}
\end{figure}

In \hyperref[fig:tau]{Fig.~\ref{fig:tau}}, we plot the median NMSE of all UEs, where the NMSE is computed with respect to their MAPs. The performance of all channel estimation schemes is shown for varying $\tau_p$, $L=8$, $K=4$, and $N=3$. It is evident that both centralized channel estimation and MACE outperform local channel estimation, thanks to their ability to leverage inter-AP signal correlation. All channel estimation schemes exhibit improved performance when $\tau_p$ increases, as expected, since more samples are used for channel estimation when $\tau_p$ increases. In \hyperref[fig:antennas]{Fig.~\ref{fig:antennas}}, we plot the median NMSE of all channel estimation schemes for varying $N$, $L=5$, $K=3$, and $\tau_p=5$. MACE and centralized channel estimation again outperform local channel estimation. Interestingly, the performance of MACE approaches that of local channel estimation as $N$ increases. The reason is that the contributions of the ASAPs decrease as $N$ increases, since the MAPs receive progressively fewer fused signals compared to $N$. For small $N$, MACE nevertheless outperforms local channel estimation. It is also observed that the simulation results deviate from the theoretical results as $N$ increases, which is attributed to the pronounced impact of estimation errors in higher dimensions.

\section{Conclusion}
In this paper, we have proposed MACE, a partially centralized channel estimation scheme that leverages inter-AP signal correlation to improve channel estimation. In MACE, each UE is assigned one MAP and a set of ASAPs. During the pilot transmission phase, the ASAPs fuse their received signals and send them to the MAP, which uses the fused signals together with its locally received signals to leverage inter-AP signal correlation for channel estimation. MACE relaxes the fronthaul and computation requirements compared to centralized channel estimation, while providing consistent performance improvements over local channel estimation.

\bibliographystyle{IEEEtran}
\bibliography{refs}

\end{document}